\documentclass{vldb}
\usepackage{graphicx,epstopdf}
\usepackage{balance}  
\usepackage{enumitem}
\usepackage{amsmath}
\usepackage{algpseudocode}
\usepackage[]{algorithm2e}
\usepackage{algpseudocode}
\usepackage[toc,page]{appendix}
\usepackage{tikz}
\usetikzlibrary{arrows,positioning,shapes.geometric}
\begin{document}


\title{Garbage Collection Techniques for Flash-Resident Page-Mapping FTLs}


%
%
%
%

 \numberofauthors{2} 
 \author{
 \alignauthor Niv Dayan\\
        \affaddr{IT University of Copenhagen}\\
        \affaddr{Copenhagen, Denmark}\\
        \email{nday@itu.dk}
 \alignauthor Philippe Bonnet\\
        \affaddr{IT University of Copenhagen}\\
        \affaddr{Copenhagen, Denmark}\\
        \email{phbo@itu.dk}
 }

\maketitle

\begin{abstract}

Storage devices based on flash memory have replaced hard disk drives (HDDs) due to their superior performance, increasing density, and lower power consumption. Unfortunately, flash memory is subject to challenging idiosyncrasies like erase-before-write and limited block lifetime. These constraints are handled by a flash translation layer (FTL), which performs out-of-place updates, wear-leveling and garbage-collection behind the scene, while offering the application a virtualization of the physical address space. 

A class of relevant FTLs employ a flash-resident page-associative mapping table from logical to physical addresses, with a smaller RAM-resident cache for frequently mapped entries. In this paper, we address the problem of performing garbage-collection under such FTLs. We observe two problems. Firstly, maintaining the metadata needed to perform garbage-collection under these schemes is problematic, because at write-time we do not necessarily know the physical address of the before-image. Secondly, the size of this metadata must remain small, because it makes RAM unavailable for caching frequently accessed entries. We propose two complementary techniques, called Lazy Gecko and Logarithmic Gecko, which address these issues. Lazy Gecko works well when RAM is plentiful enough to store the GC metadata. Logarithmic Gecko works well when RAM isn't plentiful and efficiently stores the GC metadata in flash. Thus, these techniques are applicable to a wide range of flash devices with varying amounts of embedded RAM.

\end{abstract}

\section{Introduction}

In recent years, usage of storage devices based on NAND flash memory such as eMMCs (embedded multimedia card) and SDDs (solid state drives) has been increasing at an exponential rate. The benefits of NAND flash include superior performance relative to HDDs, shock-resistance, a gradually increasing storage density, and lower power consumption. 

However, flash memory is subject to challenging idiosyncrasies \cite{tradeoff}. In particular, flash is organized into erase-blocks. Data can be written sequentially within an erase block, but any update to this data, however small, must be preceded by an erase operation, which is expensive and works at a block granularity. Moreover, an erase-block becomes increasingly prone to data errors as a function of the number of erases it endures. 

SSDs use a software layer called the flash translation layer (FTL) to manage these characteristics. An FTL's main job is to implement out-of-place updates to avoid having to erase and rewrite an entire block for every small data update. The FTL does this by providing a mapping scheme from logical to physical addresses, and a garbage-collection mechanism to reclaim invalid space. The FTL also performs wear-leveling to ensure blocks across the device wear out at the same rate. The FTL typically stores metadata in RAM\footnote{SRAM is often used rather than DRAM due to its high performance and low power consumption \cite{survey}} module embedded within the flash device. 

The simplest implementation of a mapping scheme is a RAM-resident array that maps logical to physical addresses \cite{envy, assar}. The problem is that the size of this table is often too large to fit into the RAM typically embedded in a flash device. Indeed, for consumer SSDs and portable electronics, the cost and size of a flash device is a central priority \cite{consumer}. Manufacturers typically  strive to reduce this cost by providing as little RAM as possible, especially since the cost per byte of flash memory is scaling faster than the cost per byte for SRAM \cite{survey, dftl}.  

Over the past decade, numerous RAM-efficient mapping schemes have been proposed, as captured by a recent survey \cite{survey}. Flash-resident page-associative schemes, particularly DFTL and LazyFTL, have been acknowledged as the most efficient among the schemes \cite{dftl, LazyFTL, survey}. Such schemes store a page-associative mapping table in flash, and cache frequently accessed entries in the available RAM. 

However, the design of such schemes as described in the literature is incomplete, and the missing piece is garbage-collection. The two challenges we identify and tackle in this paper concern the metadata needed by the garbage-collection module to select a victim block and to infer which pages on the victim block are still valid. We show that maintaining this metadata is a challenge because at write-time, we do not necessarily know the physical address of the before-image. We also identify the size of the metadata required for garbage-collection as an issue for flash devices that have little RAM. 

We propose two complementary schemes that address these problems. The first is called Lazy Gecko, standing for \textbf{Lazy} \textbf{G}arbage-\textbf{C}ollector. This scheme uses a flash-resident reverse map and a RAM-resident bitmap to enable victim-selection and live-page-identification. This scheme is feasible for SSDs with a moderate amount of embedded RAM. The second scheme, called Logarithmic Gecko, is designed for flash devices with very little RAM, such as eMMCs. Logarithmic Gecko is similar to Lazy Gecko, but it stores most of its metadata on flash. This requires a moderate amount of internal IOs to maintain and access, but it leaves a significantly lower RAM footprint. 

Our contributions in this paper is as follows:

\begin{itemize}
\item We introduce the problem of maintaining the metadata needed to enable garbage-collection on flash-resident page-mapping FTLs.
\item We propose two complementary techniques for solving the garbage-collection metadata problem. These techniques are suitable for a wide range of flash devices with varying amounts of RAM. 
\item We evaluate these techniques using simulation and demonstrate their impact on write-amplification and read-amplification.
\end{itemize}

\section{Related Work} \label{sec:related_work}

\subsection{Flash Devices} \label{sec:flash}

Flash devices store data in NAND chips, each of which is organized into independent arrays of memory cells. Each cell accommodates 1, 2 or 3 bits (SLC, MLC or TLC). An array is a flash block, and a row within the array is a flash page. A flash page can typically store 4-16 KB and a block typically contains 128-512 pages respectively. 

Flash devices are subject to challenging idiosyncrasies.  An erase operation must precede an update to a page, and an erase has the granularity of a block. Moreover, blocks have a limited lifetime in terms of erases \cite{tradeoff}.

Unfortunately, as the density of flash devices is increasing, their reliability and the lifetime of blocks are decreasing \cite{bleak}. Bit-shifts may occur both when writing and reading a page due to electrical side-effects. To mitigate this risk, two additional constraints on writes are generally applied for MLC and TLC devices. Writes have a minimal granularity of a flash page, and writes must take place sequentially within a block \cite{truth}. 

Each flash page is given an adjacent spare out-of-bound (OOB) area for storing metadata about a page.  It is typically smaller than the page itself by a factor of 32 and contains the logical address of a page as well as error-correction codes. 

\subsection{Log-Based Hybrid FTL Schemes}

In log-based hybrid schemes \cite{survey}, the blocks in the SSD are divided into two types: data blocks and log blocks. Logical pages with adjacent addresses are stored in the order of their addresses on data blocks. There is a mapping table in RAM from logical blocks to physical blocks. A page can be looked up by determining its block from the mapping and adding its modulo offset within the block. The remaining log blocks are used to buffer updates. A log block is page-associative, meaning page updates on it can store updates in any order. A page-mapping for all log blocks is stored in RAM. 

When space for incoming updates becomes limited, a log block is selected and merged with the data block that have updates on it. The more data blocks that have a page update on a log block, the more the cost of the merge increases, as more data blocks must be erased and rewritten. In the Set-Associative Sector Translation (SAST) scheme \cite{sast}, log blocks are set-associative. This means that the logical address space is divided into equally sized sets, and a given log block can only accept updates from pages that belong in a given set. This limits the cost of a merge by restricting the number of data blocks that can be associated with a log block. SAST is used in practice by the very recent F2FS \cite{F2FS} file system for iMMC devices. Other log-based hybrid schemes of interest are BAST \cite{BAST} and FAST \cite{FAST}, whereby log blocks are block-associative and fully-associative respectively. 

\subsection{Flash-resident page-associative schemes}

In flash-resident page-associative schemes, the mapping table is page-associative, meaning that a logical page can be written on any physical page. Since this requires a large mapping table (typically 4 bytes per flash page), this table is stored and maintained in flash. Frequently accessed mapping entries are cached in RAM \cite{survey}. 

For example, in DFTL \cite{dftl} a mapping page stores mapping entries for adjacent logical addresses. The mapping in flash is updated lazily. This means that when a page is updated, the updated mapping entry will first only exist in the RAM-resident mapping cache. Such an entry is labelled dirty. The cache uses a LRU page-replacement strategy to evict entries as it runs out of space. If a dirty entry is evicted, its corresponding mapping page in flash must be read, updated and rewritten.  

LazyFTL \cite{LazyFTL} is an interesting variation of DFTL that separates hot and cold data and strives to provide better consistency guarantees to avoid losing cached addresses at the event of power failure. 

Flash-resident page-associative schemes tend to involve significantly lower write-amplification than log-based hybrid schemes. The reason is that there is complete flexibility in where we can store a page. Thus, we avoid the potentially expensive merge operations that are inherent in log-based hybrid schemes. Instead, we can simply pick a block with few live pages, migrate these pages into other blocks with free space, and erase the block. A lower write-amplification implies better performance, device longevity and reliability. 


The challenge with page-associative schemes, however, is that they rely on high temporal locality in the data. The larger the working set in the workload is, the more cache misses and evictions occur, which lead an increase and read-amplification and write-amplification. It is therefore desirable to allocate as much of the available RAM as possible to the cache for frequently accessed entries. 

\subsection{Shortcomings of Existing Schemes}

Although flash-resident page-associative mapping schemes are the state-of-the-art in terms of performance and reliability, existing schemes have problems. In particular, DFTL does not address the problem of how to maintain metadata to determine which pages are valid and should be migrated as a part of a garbage-collection operation. 

LazyFTL does address this problem, but its design relies on an obsolete assumption the OOB area of a page can store a bit that indicates if the page is valid or invalid. This assumptions dates back to 1995 in the design of a Flash-File System \cite{ffs}, whose design assumed that a part of the OOB could be left un-programmed initially, and programmed later to indicate when the page has become invalid. Unfortunately, as we saw in Section \ref{sec:flash}, a new constraint recently emerged that pages should be programmed sequentially within a block to minimize electrical side effects. The design of LazyFTL violates this constraint. 

The problem of maintaining metadata for garbage-collection has not been addressed adequately in any work we are aware of. It is concerning that the recent F2FS file system uses the SAST FTL as opposed to a state-of-the-art page-associative schemes like DFTL and LazyFTL. The reason may be that page-associative schemes are still yet to have reached maturity. We hope that this paper would help in filling this gap. 

\section{Problem Definition}

In the context of flash-resident page-associative FTLs, any garbage-collection scheme must answer two questions: which block to reclaim next, and which pages are still valid on this victim block? As we will now see, those questions are straight-forward to answer when SRAM is abundant and the mapping table can fit into RAM. As we will later see, those questions are much more difficult to answer when RAM is scarce and the mapping table is in flash. Table \ref{tab:terms} lists terms we use throughout the paper. We refer to the logical to physical mapping table as \textit{page-mapping}.

\begin{table*}
\caption{Terms}
\centering 
\begin{tabular}{|p{0.15\linewidth}|p{0.40\linewidth}|p{0.15\linewidth}|p{0.15\linewidth}|}
\hline
Term & Description & Micron P420m & Intel 525 series  \\ \hline
$K$ & Number of blocks in the SSD & $2^{18}$ & $2^{16}$ \\ \hline
$B$ & Number of pages per block & 512 & 128 \\ \hline
$P$ & Size of a page & 16 KB & 4 KB  \\ \hline
$OP$ & Over-provisioning, measured as 1 minus the size of the logical address space over the size of the physical address space & $30\%$ & $7\%$ \\ \hline
$PBA$ & $K \cdot B$ (number of physical pages) &   &    \\ \hline
$LBA$ & $K \cdot B \cdot (1 - OP)$ (number of logical pages) & & \\ \hline
$a$ & Page or block address. Assume $4$ bytes. &  &  \\ \hline
$R$ & Size ratio between adjacent levels in Logarithmic Gecko's LSM-tree &  &  \\ \hline
$L$ & Maximum number of levels in the LSM-tree &  &  \\ \hline
\end{tabular}
\label{tab:terms}
\end{table*}


\subsection{Abundant RAM}

Assuming RAM is plentiful enough to store the entire page-mapping, let us survey a few techniques for performing victim-selection and live-page-identification.

\subsubsection{Victim Selection}

\textbf{\textit{Greedy}}: A simple method for victim selection is to maintain a mapping from block id to the number of valid pages in each block. 
To select a GC victim, we scan this mapping and choose the block with the least number of live pages. To maintain this mapping, we must know the physical address of the before-image of every update. We use the address of the before image to decrement the counter for the block in which the before-image resides. 

\textbf{\textit{LRU}}: A different technique for victim-selection is least recently used (LRU), which selects the block that was erased the longest time ago. The rationale is that this block is likely to contain the least number of live pages. Typically, this requires maintaining a queue of blocks. A block is inserted into the queue when it is written, and a victim is selected by popping the queue. The issue of the LRU scheme is that it may involve more migrations than the greedy scheme since we have no guarantee we have actually chosen the block with the least number of live pages. 

\textbf{\textit{Window-greedy}}: A compromise between the LRU and greedy policies is window-greedy. It implements a block queue like the LRU policy and applies the greedy policy only to the front X blocks in the queue. This allows avoiding the potentially CPU-expensive scan of the greedy algorithm, and to increase the chance of finding a block with few live pages relative to the LRU policy.


Note that some methods also choose a victim based on age \cite{ffs}. Such methods essentially integrate the wear-levelling and garbage-collection schemes. In this work we just concentrate on garbage-collection. Some works also separate pages into groups based on update frequency and perform garbage-collection independently within each group \cite{stoica}. This is also outside of the current scope. 

\subsubsection{Live Page Identification} \label{sec:live_page_iden}

Once a victim has been chosen, we need to check which pages in it are still valid. Three techniques are possible. 

\textbf{\textit{page-mapping scan}}: We can scan the entire \textit{page-mapping} to find all live pages that are on the target block. However, this scan may become a CPU bottleneck.

\textbf{\textit{page-validity-bitmap (PVB)}}: A less CPU-intensive alternative is to use a \textit{page-liveness-bitmap}, which tracks which pages in the SSD are valid and which are invalid. Pages clustered based on which block they are on. To maintain this map, we must know the physical location of the before-image of each write to shift the corresponding bit in the bitmap. Note that if we use the greedy policy for victim selection, the PVB can be used to keep track of the number of live pages in a block by taking the Hamming weight of the bits associated with a given block.

\textbf{\textit{flash-reverse-mapping}}: Yet another alternative is to store a \textit{flash-reverse-mapping} in the out-of-bound component of each block. This mapping indicates which logical pages are written in each of the physical pages on the block. It is updated when the block is written. In order to identify which pages are valid, we read this map before starting a GC operation. We look up each logical addresses in the page-mapping table and check if the physical address still corresponds to the block we are targeting. If so, then the page is valid.

Note that the above three techniques assume that \textit{page-mapping} is in RAM. In the next section, we examine the problems that arise when most of \textit{page-mapping} is stored in flash. 

\subsection{Scarce RAM}

When RAM is scarce and most of \textit{page-mapping} is stored in flash, challenges arise. The essential problem in all cases is that when a write arrives, we may not know the physical location of its before-image. Thus, we cannot keep the metadata up-to-date. Let us re-examine the policies from the last subsection in this context. 

\subsubsection{Victim Selection} 

\textbf{\textit{Greedy and window-greedy}}: In order to keep track of the number of live pages per block, we must know the physical location of the before-image to decrement the appropriate block counter. However, if the page's mapping entry is not cached, the address of the before image is unavailable, and we don't know which counter to decrement. We can look it up in \textit{page-mapping} in flash using a read IO, but doing so for each write can severely increase read-amplification.

\textbf{\textit{LRU}}: The LRU policy is unaffected by moving \textit{page-mapping} to flash as it does not rely on accessing it. We leverage this later. 

\subsubsection{Live Page Identification} 

\textbf{\textit{page-mapping scan}}: Scanning \textit{page-mapping} to determine which pages are still valid in each target block would cripple performance, since \textit{page-mapping} is in flash and comprises thousands of flash pages. 

\textbf{\textit{page bitmap}}: Maintaining a RAM-resident bitmap that indicates which page is live is problematic. The reason is that when we update a page, we don't know the physical address of the before-image if it is not pre-cached in \textit{page-mapping cache}. Thus, we cannot shift the appropriate bit in the bitmap to indicate the page is now invalid. We can access \textit{page-mapping} to find the before-image, but doing so for each write is going to significantly increase read-amplification. 

\textbf{\textit{flash-reverse-mapping}}: This policy, which involves reading a reverse-mapping from the OOB part of the target block, is also impractical. The reason is that for each entry in the reverse mapping, we need to access \textit{page-mapping} to determine whether the logical page is still on the same block. Typically, most of these addresses will not be cached, and so \textit{page-mapping} may need to be accessed up to B times for each GC operation. This greatly increases read-amplification. 

\subsection{Problem Summary}

\textbf{The Bookkeeping Maintenance Problem}: The metadata needed for victim-selection and live-page-identification requires being maintained for every page update with information about the physical address of the before-image. However, the mapping entry with the before-image may not be cached. The naive solution is to access page-mapping using read IOs to find the before-image's physical address, but doing so too much may lead to unacceptable read-amplification.

\textbf{The Scarce RAM Problem}: Another problem is that this metadata  may consume a substantial amount of RAM. The RAM consumed by this metadata becomes unavailable for caching frequently accessed entries, which degrades performance. It also limits the ability of SSDs to scale, as RAM is more expensive than flash \cite{survey}. 

In the next two sections, we propose two novel garbage-collection algorithms, namely Lazy Gecko and Logarithmic Gecko. The former only addresses problem 1. The latter is an extension of the former that also addresses problem 2. 

\section{System Model} \label{sec:model}

Before describing the two schemes, let us outline some assumptions about the underlying system. We consider an SSD whose architecture is captured by the terms in table \ref{tab:terms}. We assume that this SSD does not have sufficient RAM for storing an entire page-mapping in flash, and that a flash-resident page-associative FTL is used to store page-mapping in flash. For concreteness and for the experimental evaluation later, we assume the mapping scheme is DFTL as described in \cite{dftl}. However, the techniques introduced in this paper are in principle also applicable to LazyFTL, or any other flash-resident page-associative scheme. 

DFTL stores the mapping table in translation pages in flash. These translation pages occupy separate blocks from user data. The reason for this is that translation pages are updated much more frequently, and it is considered beneficial for performance to separate pages of different update frequencies on different flash blocks \cite{stoica, Desnoyers, envy}. There is a pool of free blocks, and one active translation block and one active data block on which translation pages and data pages are written respectively. When an active block of either groups runs out of space, a new one is requested from the pool of free blocks. When the pool of free blocks is nearly empty, the garbage-collection mechanism is invoked. 

A RAM-based table called the Global Mapping Directory (GMD) stores the locations of translation pages in flash. Finally, DFTL stores frequently accessed mapping entries in a RAM-based table called the Cached Mapping Table (CMT). We refer to an entry in the CMT as "dirty" if it is not synchronized with the corresponding mapping entry in flash.  

\section{Lazy Gecko} \label{sec:lazy_map}

Here we introduce Lazy Gecko, which stands for \textbf{Lazy} Bookkeeping \textbf{G}arbage-\textbf{C}ollector. Lazy Gecko combines a few of the techniques described in the problem definition to solve the bookkeeping maintenance problem.

\subsection{Data Structures} \label{sec:lazy-data}

In terms of data structures, Lazy Gecko uses a RAM-resident bitmap called the Page Validity Bitmap (PVB). There is one bit for each page in the system. A bit is set to 1 if the page is invalid, and 0 otherwise. It is updated using the following trivial algorithm \ref{algo:invalidate1}, which we override later in the design of Logarithmic Gecko. 

\begin{algorithm}
\TitleOfAlgo{invalidate()}
\LinesNumbered
 \KwIn{physical$\_$address pa}
    PVB[pa] = 1    
   	\caption{marks a physical page as invalid.}
   	\label{algo:invalidate1}
\end{algorithm}

Lazy Gecko also maintains a flash-resident \textit{Reverse-Map}, a mapping from physical pages to the logical pages that were last written on them. The reverse-map is stored on flash pages rather than the OOB component of flash pages. This map occupies a small percentage of flash (approximately $0.02\%$ in the devices in table \ref{tab:terms}). It is arranged such that the mapping entries of all physical pages from the same flash block are stored on the same flash translation page. A RAM-resident Global-Reverse-Mapping-Directory is used to keep track of the whereabouts of the flash pages in this map. Translation pages belonging to the reverse map are stored on separate flash blocks, which we call \textit{Reverse Blocks}. The reverse map is updated as follows. Whenever a flash block containing user data is written, its corresponding reverse translation page is read, updated to reflect the new logical addresses written to the block, and written to flash on a Reverse Block with free space. Maintaining this map entails a modest overhead of one read and one write IO per garbage-collection operation. 

Finally, we add one bit to each entry in the cached mapping table called the "synch flag". This flag is set to 0 if a logical address in the cache has some before-image whose physical address has still not been set to invalid in PVB, and 1 otherwise.

\subsection{Operations}

The crux of Lazy Gecko is how to maintain the PVB up-to-date in order to allow performing victim-selection and live-page-identification. This is done by adding some logic to the following operations.

\subsubsection{Application Write}

When an application page update arrives, we follow algorithm \ref{algo:1}. This algorithm checks if the current mapping entry for the logical address is in CMT. If not, we insert it into the CMT, and the synchronization flag is set to false. If it is cached, we invoke algorithm \ref{algo:invalidate1} to invalidate the page's former physical address. However, we do not set the synchronization flag to true, as there may be another physical page on which the logical page once resided that is still marked as valid in the PVB. Lastly, we execute the write and update the cached entry with the new physical address of the logical page. 

Note that this procedure creates false-positives in the PVB. In other words, there may be pages in the bitmap marked as valid that are actually invalid. We show how to resolve these false positives later.

\begin{algorithm}
\TitleOfAlgo{Handle Write}
\LinesNumbered
 \KwIn{page$\_$write}
   la = page$\_$write.logical$\_$address\;
   \eIf{cache.contains(la)}{
        pa = cache[la].physical$\_$address\;
        invalidate(pa)\;
  	} {
  	    cache.insert(la)\;
  	    cache[la].synch$\_$flag = false\;
  	}
    cache[la].physical$\_$address = ssd.write(page$\_$write)\;
    cache[la].dirty = true \;
     \caption{ Handles an application page write. }
 \label{algo:1}
\end{algorithm}

\subsubsection{Application Read}

We handle an application read as follows. If the mapping address is cached, we simply execute the read. If it is not cached, we issue a read IO to the appropriate flash translation page. We then insert the mapping entry into the CMT with the synchronization flag set to true. We omit an algorithm listing for this operation due to space constraints.

\subsubsection{Translation Page Read} \label{sec:translation_page_read}

We resolve false positives in PVB by piggybacking some logic onto routine operations that take place in the background all the time, namely (1) cache misses, (2) cache evictions, and (3) GC migrations targeting translation pages. Whenever one of these operations takes place, we invoke algorithm \ref{algo:2}, which iterates through all the mapping entries in the translation page. If any of them is in CMT with the synchronization flag marked as false, it means that the physical page on which the logical page was at some point written has not been marked as invalid in the PVB. We correct this by setting the corresponding bit in PVB to 1. We also set the synch flag for the cached entry to true, because at this point there can be no other before-images for the logical address that are still unsynchronized. The reason is that algorithm \ref{algo:2} is always called as a result of a page eviction, so a page is always synchronized when it is 
evicted. 

\begin{algorithm}

\TitleOfAlgo{Lazy Updates}
\LinesNumbered
 \KwIn{mapping$\_$page}
  \ForAll{entries in mapping$\_$page} {
    la = entry.logi$\_$addr\;
    pa = entry.phys$\_$addr\;
  	\If{cache.contains(la) and !cache[la].synch$\_$flag}{
        invalidate(pa)\;
        cache[la].synch$\_$flag = true\;
  	}
  }
 \caption{Detects and resolves false positives.}
 \label{algo:2}
\end{algorithm}

Note that this algorithm \ref{algo:2} cannot be a bottleneck as the number of entries in a translation page is typically between $2^{10}$ and $2^{12}$. Iterating through an array of this size takes hundreds of nanoseconds, whereas the cost of flash operations is in the order of tens to hundreds of microseconds. 

\subsubsection{Garbage-Collection}

In terms of victim-selection, Lazy Gecko is compatible with the Greedy and Window-Greedy, as the PVB can be scanned to select the block with the fewest live pages. The number of live pages in a block is given by taking the Hamming Weight (number of non-zero bits) of its bitmap. 

Live-page-identification works by referring to PVB. However, when a block is chosen for garbage-collection, false positives may still exist in the PVB. To resolve these, we use the following key insight: if a physical page marked as valid PVB is actually invalid, a dirty mapping entry for the logical page which was last written on it must still be the Cached Mapping Table. The reason for this is that when a dirty mapping entry is evicted from the cached mapping table, the PVB is updated. Thus, if PVB is not up-to-date, it means a page eviction never took place, and the logical entry with its current physical page address are cached. 

We exploit this insight as follows in algorithm \ref{alg:false}, which resolves any remaining false positives. We read the reverse translation page from Reverse Mapping corresponding to the victim block, and iterate through the logical addresses that were last written to this block. If any of them is in the cached mapping table with the synchronization flag set to false, we know that the physical page in the victim is in fact invalid. This is all we need to complete the live-page identification. 

\begin{algorithm}

\TitleOfAlgo{Garbage Collection}
\LinesNumbered
 \KwIn{victim block b}
  original$\_$block$\_$mapping = ssd.read$\_$reverse$\_$mapping(b)
  \ForAll{entries in original$\_$block$\_$mapping} {
    la = entry.logical$\_$address\;
  	\If{cache.contains(la) and !cache[la].synch$\_$flag}{
  	    pa = entry.physical$\_$address\;
        invalidate(pa)
  }
}
\caption{}
\label{alg:false}
   
\end{algorithm}

When a block is erased, we reset the bits corresponding to its physical pages in the PVB. We also update \textit{Reverse Mapping}, as described in subsection \ref{sec:lazy-data}.





\subsection{Reflection}

We showed that in order to resolve the garbage-collection metadata maintenance problem, Lazy Gecko maintains a page-validity-bitmap (PVB) in RAM and stores a reverse map in flash. The only IO overhead is introduced is 1 flash read and 1 flash write to update the reverse-mapping for each garbage-collection operation. This overhead is relatively modest. 

The main problem of Lazy Gecko is that the amount of RAM needed is proportional to the number of pages in the SSD, as 1 bit is needed for each page. This may be an issue for SSDs with little RAM. For instance, an SSD of the same dimensions as the Micron P420m in table \ref{tab:terms} requires 16 megabytes for the bitmap. This is a hard limit. An SSD with less RAM than this cannot use Lazy Gecko, and can therefore not use a flash-resident page-associative scheme. And even if the SSD has enough RAM to store the PVB, the RAM consumed by PVB is unavailable for storing frequently accessed entries in CMT, thereby degrading performance. These problems are addressed in the next section. 

\section{Logarithmic Gecko}

We now introduce Logarithmic Gecko, which stands for \textbf{Logarithmic} \textbf{G}arbage-\textbf{C}ollector. It is very similar to Lazy Gecko, the main difference being that in Logarithmic Gecko the Page Validity Map is stored in flash to save RAM. This reduces the RAM footprint by as much as $97\%$ relative to Lazy Gecko for flash devices that are on the market today. Logarithmic Gecko requires a modest number of IOs to maintain the flash-resident PVB.  

Logarithmic Gecko uses a \textit{Reverse Map} and a Page Validity Bitmap (PVB) similarly to Lazy Gecko. It is identical to Lazy Gecko in terms of how PVB is updated lazily, and how we resolve false positives using the Reverse Map. The difference is that Logarithmic Gecko stores PVB in flash as an LSM-tree \cite{lsm-tree}. As an overview, the PVB is structured as a series of "sorted runs" of exponentially increasing sizes in flash. Each sorted run contains a sorted mapping from block ids to bitmaps. A bitmap has B bits, one for each physical page in the block, which indicate whether they are valid or not. The first sorted run is in a RAM-based buffer, and updates are made to it as discussed in Section \ref{sec:buffer}. When this buffer fills up, it is flushed to flash, and then a merge procedure may commence which merges two or more sorted runs, as discussed in \ref{sec:merging}. To perform live-page-identification for a given block, we search for its id in all the sorted runs and apply the bitwise "or" operator product of all its bitmaps, as described in Section \ref{sec:query}. Victim-selection is discussed in Section \ref{sec:victim_sel}, and further possible optimisations are discussed in Section \ref{sec:optim}.

We emphasize that Lazy and Logarithmic Gecko are identical in terms of the logic of live-page-identification. The crux of this section is about how to keep PVB in flash. 

\subsection{Buffer Management} \label{sec:buffer}

Logarithmic Gecko has one buffer in RAM the size of one flash page. This buffer contains a sorted mapping from block ids to \textit{Gecko Entries}. A Gecko entry consists of two fields: (1) a bitmap of size B, where the bit at offset $i$ corresponding to whether the physical page at offset $i$ in the block is invalid, and (2) one additional bit called the "erase flag", which is used to indicate that a block has been erased. The erase flag is used for merging (see next subsection).

Logarithmic Gecko handles application reads and writes just as Lazy Gecko does. However, the \textit{invalidate} procedure from algorithm \ref{algo:invalidate1} is overloaded, since PVB is no longer a simple RAM-based bitmap. Instead, algorithm \ref{algo:invalidate2} is invoked. This algorithm is given a physical address of a page that is no longer valid. It firstly checks if an entry for the block id of the before-image is in the buffer. If not, it adds an entry with the block id as the key and a blank gecko entry (the bits in the bitmap and the erase flag are all set to 0). The bit in the bitmap that corresponds to the page that has been invalidated is then set to 1.

\begin{algorithm}
\TitleOfAlgo{invalidate()}
\LinesNumbered
 \KwIn{physical$\_$address pa}
    block$\_$id = pa.block$\_$id\;
    page$\_$offset = pa.page$\_$offset\;
    \If{!buffer.contains(block$\_$id)} {
   		buffer.insert(block$\_$id)\;
        buffer[block$\_$id].bitmap = blank bitmap\;
        buffer[block$\_$id].erase$\_$flag = false\;
   	} 
   	buffer[block$\_$id].bitmap[page$\_$offset] = 1 \;
   	\If{buffer is full} {
   		flush(buffer) \;
   	} 	    
   	\caption{}
   	\label{algo:invalidate2}
\end{algorithm}

When the buffer is filled up, its contents are flushed to flash and it is cleared so new entries can be written on it. The flush may trigger a processes that merges one or more of the flash-resident sorted runs. This is described in the next section. Note that all flash pages that comprise the LSM-tree are allocated on separate flash blocks that we refer to as Gecko Blocks.

Resolving false positives in PVB is done using algorithm \ref{algo:2} in Section \ref{sec:translation_page_read}. The only difference is that within algorithm \ref{algo:2}, the new version of the \textit{invalidate} method is invoked instead of algorithm \ref{algo:invalidate1}.  

When a regular data block is erased and written with new entries, the procedure in algorithm \ref{algo:erase} is invoked. If there is already an entry corresponding to the block in the buffer, its erase flag is set to 1. Otherwise, an entry is inserted with a blank bitmap and the erase flag set to 1. 

\begin{algorithm}
\TitleOfAlgo{Handle$\_$Block$\_$Rewritten}
\LinesNumbered
 \KwIn{block$\_$id}
  	\If{buffer.contains(block$\_$id)}{
        buffer.insert(block$\_$id)\;
        buffer[block$\_$id].bitmap = blank bitmap\;
  	} 
  	buffer[block$\_$id].erase$\_$flag = true\;
  	\caption{}
  	\label{algo:erase}
\end{algorithm}

\subsection{Merging} \label{sec:merging}

In the last subsection, we saw that when the buffer fills up, it is flushed to a gecko block on flash. We call the flushed page a "sorted run" of size 1, and we say that it is at level 1 of the LSM-tree. (We consider the RAM-based buffer to be at level 0.)

As any LSM-tree, the Logarithmic Gecko tree contains multiple levels, and we denote the $n^\mathrm{th}$ level as $L_n$. There is typically either 0 or 1 sorted run per level. The LSM-tree has a tuning parameter T, which determines the size ratio of sorted runs in any two adjacent levels of the tree. A sorted run at level $i$ contains between $T^{i-1}$ and $T^{i}-1$ flash pages. We discuss the impact of the parameter T on performance in Section \ref{sec:analysis}.

If there is more than one sorted run at level i, a merge procedure is triggered. This procedure allocates two input buffers and one output buffer in RAM. It stores the resulting run in level $i$ or $i+1$ depending on how many pages it has, and disposes of the original runs. Thus, a merge may be invoked after a flush, and it may continue recursively based on the state of the tree.  

During a merge, if two sorted runs contain entries with the same block id, the following rule in algorithm \ref{algo:merge} is followed. We assume that entry1 is from the more recently created tree. The erase flag is used to discard any entries from before the last time this block was erased. Otherwise, the bitwise \textit{or} operation is used to merge the bitmaps.

\begin{algorithm}
\algnewcommand{\And}{\textbf{and}}

\TitleOfAlgo{merge entries}
\LinesNumbered
 \KwIn{entry1, entry2}
  	\eIf{entry1.erase$\_$flag == true}{
        \Return entry1 \;
  	} {
  	  entry1.bitmap = entry1.bitmap \textbf{or} entry2.bitmap
  	  
      entry1.erase$\_$flag = entry2.erase$\_$flag 
  	  
  	  \Return entry1 \;
  	}
  	\caption{}
  	\label{algo:merge}
\end{algorithm}

We maintain a RAM-based data structure called the Logarithmic Gecko Mapping Directory (LGMD), which keeps track of the whereabouts of all flash pages belonging to the LSM-tree. For each entry, we also include the values of the first key within the page.  

\subsection{Live-Page-Identification} \label{sec:query}

We can use the LSM-tree to perform live-page-identification as follows. When a victim block has been selected, we search the sorted runs one by one starting at the lowest level either until we have searched all of them, or until we encounter an entry whose erase flag is set to true. Searching a level requires exactly 1 IO because a run is sorted and we can use LGMD to infer the only candidate page in the run that can store the key, since LGMD stores the starting key of each page in the run. 

Once we have finished the search, we perform a bitwise \textit{or} operation on all of the bitmaps we found. We then invoke algorithm \ref{alg:false} to resolve any remaining false positives in the resulting bitmap. This gives us an up-to-date image of which pages are currently live in the block. 

\subsection{Victim-Selection} \label{sec:victim_sel}

Recall from Section \ref{sec:model} that in our system there is a pool of free blocks. When this pool is nearly empty, the garbage-collection mechanism is invoked. We now discuss how this mechanism performs victim-selection. 

There are four types of blocks in Logarithmic Gecko: translation blocks (storing the conventional logical to physical page mapping), reverse Blocks (storing the reverse page mapping), Gecko blocks (storing the LSM-tree), and data blocks, (storing user data). For convenience, we refer to translation, reverse and Gecko blocks collectively as internal blocks. Note that internal blocks occupy less than $1\%$ of all blocks on the system. 

For the purpose of victim-selection, Logarithmic Gecko treats data blocks and internal blocks differently. For internal blocks, page validity bitmaps are stored in RAM. They are maintained lazily using Lazy Gecko. 

For data blocks, we apply the LRU policy using a simple queue, called the Data Block Queue (DBQ). When a block is erased and rewritten, its id is inserted into the queue. For victim-selection, we pop the top of the queue and search the Logarithmic Gecko LSM-tree with the block id as the key as in Section \ref{sec:query}. This gives us the page validity map for the candidate block, and we cache it in RAM.

The greedy strategy is ultimately applied to pick the block with the least number of live page from among the internal blocks and the cached data block candidate. Note if the cached data block candidate contains cold data, it may never be picked. We apply a rule whereby the candidate is discarded if it is not picked after 3 victim-selection processes. Note that if the cached data block is selected, we immediately start searching for the next data block candidate (by popping the DBQ and looking up the block id in the LSM-tree).

\subsection{Further Possible Optimisations} \label{sec:optim}



\subsubsection{Compression}

We can reduce the IO overhead of merging the tree by compressing the bitmaps. Our key insight here is that when the LSM-tree buffer is flushed, most of the bitmaps within it, the vast majority of the entries only have 1 bit set to 1. This is because the number entries that fits into the buffer is vastly smaller than the number of blocks in the system. We can exploit this by not storing a full bitmap (e.g. 16 bytes for a 128 page Block), but only storing the offsets of the pages that are invalid for the first few levels of the tree (e.g. 4 bytes). This allows us to store four times  more entries in the buffer before we need to flush it, thereby significantly reducing the number of LSM-tree 

\subsubsection{Multi-way Merge} \label{sec:multiway}

As mentioned in Section \ref{sec:merging}, the merging of adjacent levels may continue recursively, as long as the result of one merge leads to the existence of more than one runs in the next bigger level. Note that this is wasteful in terms of IOs, as we continue merging the same data from the lower runs several times. We can reduce the number of IOs by pro-actively determining how far many levels the recursive merge will have encompassed, and instead performing a multi-way sort merge. The new criteria for a run at level $i$ to participate in a merge is if: (1) it is not already participating in another merge, (2) there is at least one run at level $i-1$ participating in this merge, and (3) all the runs at level $i-1$ which are participating in the merge have a combined size of at least $s = (T^i - T^{i-1})$. These rules are simple. The only downside is that more input buffers are needed in RAM to perform the multi-way merge. If L is the number of levels in the tree, then we need at most L buffers. 

\subsubsection{Flash-Resident Queue} \label{sec:queue}

Finally, the DBQ may occupy a substantial amount of RAM relative to the other RAM-resident data structures in Logarithmic Gecko. For instance, for a device with the dimensions of the Micron P420 in table \ref{tab:terms}, the DBQ takes up 1 MB of RAM. Luckily, it is very easy to store most of this queue in flash. We use an input buffer into which blocks are appended when they are erased. When it runs out of free space, it is flushed to flash. A RAM-based structure called the \textit{Queue Directory} keeps track of which flash pages belong to the DBQ. There is also an input buffer which contains the block ids that were least recently written. Block ids can be popped from this input buffer as candidates for garbage-collection. When the input buffer runs out of space, we use the DBQ to read the next queue page. The IO cost of this technique is negligible, as only 1 read IO and 1 write IO are needed for every $P \ a$ block rewritten (4096 for the Micron P420m). 

\section{Analysis} \label{sec:analysis}

\begin{table*}
\caption{Ram-resident data structures for DFTL with Logarithmic Gecko}
\centering 
\begin{tabular}{|p{0.16\linewidth}|p{0.26\linewidth}|p{0.21\linewidth}|p{0.10\linewidth}|p{0.10\linewidth}|}
\hline
Scheme & data structure & Size (bytes) & Micron P420m & Intel 525 series \\ \hline

 & Global Mapping Directory & $a (LBA / (P / a))$ & 90 KB & 22 KB \\ \cline{2-5}

DFTL with & Reverse Mapping Directory & $a (PBA / (P / a))$ & 128 KB & 32 KB \\ \cline{2-5}

Lazy Gecko & Page Validity Bitmap & $ (K \cdot B) / 8 $ & 16 MB & 1 MB \\ \cline{2-5}

 & \textbf{total} & & $\approx$ \textbf{16.5 MB} & $\approx$ \textbf{1.1 MB} \\ \hline 

 & Global Mapping Directory & $a (LBA / (P / a))$ & 90 KB & 22 KB \\ \cline{2-5}

DFTL with & Reverse Mapping Directory & $a (PBA / (P / a))$ & 128 KB & 32 KB \\ \cline{2-5}

Logarithmic Gecko & Gecko Mapping Directory  &  $ 2 a (2 \cdot (K / (P / (a + (B / 8)))))$ & 8.5 KB & 2.5 KB \\ \cline{2-5}

 & Queue Directory & $2a \cdot ((K \cdot a) / P)$ & 512 B & 512 B \\ \cline{2-5} 

 & Cached bitmaps & $  $ & 15 KB  & 4 KB \\ \cline{2-5} 

 & page buffers & $P \cdot (4 + L) $ & 241 KB & 53 KB \\ \cline{2-5} 

 & \textbf{total} & & $\approx$ \textbf{482 KB} & $\approx$ \textbf{112 KB} \\ \hline
 
  & \textbf{Ratio} & total logarithmic / total lazy & $\approx$ \textbf{3$\%$} & $\approx$ \textbf{11$\%$} \\ \hline
\end{tabular}
\label{tab:RAM}
\end{table*}

\renewcommand{\arraystretch}{1.3}
\begin{table*}[ht]
\caption{Comparison of overheads for different garbage-collection techniques}
\centering
\begin{tabular}{|p{0.20\linewidth}|p{0.14\linewidth}p{0.14\linewidth}|p{0.14\linewidth}p{0.14\linewidth}|}
\hline
& \multicolumn{2}{|c|}{\textbf{overheads for a write IO}} & \multicolumn{2}{|c|}{\textbf{overheads for a GC operation}}  \\ 
\textbf{technique} & flash read & flash writes & flash read & flash writes   \\
\hline
Lazy Gecko & \multicolumn{1}{c}{0} & \multicolumn{1}{c|}{0} & \multicolumn{1}{c}{1} & \multicolumn{1}{c|}{1}   \\
Logarithmic Gecko & \multicolumn{1}{c}{0} & \multicolumn{1}{c|}{O($\frac{T \cdot B}{P} log_T(\frac{K \cdot B}{P}))$}  & \multicolumn{1}{c}{1 + $O(log_T(\frac{K \cdot B}{P}))$} & \multicolumn{1}{c|}{1}    \\
\hline
\end{tabular}
\label{tab:IO}
\end{table*}

Table \ref{tab:RAM} shows a breakdown of the minimum amount of RAM needed by the different RAM-based data structures in Lazy and Logarithmic Gecko for our two example flash devices. The formulas and figures are derived using the the terms and values in table \ref{tab:terms}. The CMT is not listed, because it does not require a strict minimum amount of RAM. It is assumed that any leftover RAM in the system is allocated to the CMT. 

Table \ref{tab:RAM} shows that Lazy Gecko's RAM consumption for both devices is in the order of several megabytes. The dominant RAM occupant is the PVB. The other data structures, the GMD and the RMD, which store the whereabouts of translation pages for the global mapping table and the reverse mapping table, are common to Lazy Gecko and Logarithmic Gecko. 

Logarithmic Gecko requires significantly less RAM than Lazy Gecko because it stores the PVB in flash. However, Logarithmic Gecko stores several other RAM-based data structures to support the flash-based LSM-tree and queue. The Logarithmic Gecko Mapping directory (Section \ref{sec:merging}) and the queue mapping directory (section \ref{sec:queue}) keep track of the whereabouts of all flash pages containing pages belonging to the LSM-tree or to the data block queue. These mappings are small because the number of flash pages they keep track of is relatively small. As we saw in Section \ref{sec:victim_sel}, Logarithmic Gecko uses several cached bitmaps for blocks that host translation pages and LSM-tree pages, but since there are few such blocks, these bitmaps consume little RAM. We omit the calculation for the RAM-consumption of these bitmaps from the table because it is cumbersome. Finally, logarithmic Gecko uses significantly more page buffers than Lazy Gecko to support the multi-way merge (L+1 buffers), the data block queue (one input and one output buffer) buffers, and the LSM-tree (one input buffer). 

Interestingly, the relative amount of RAM saved is much higher for the larger Micron device. The reason is that the larger device has far more flash pages. The PVB grows in proportion to the number of pages, but all the other structures grow at a slower rate. Thus, as the number of pages in an SSD increases, the more relative saving we get in RAM due to logarithmic Gecko. 

Is the magnitude of RAM-saving by Logarithmic Gecko significant? To answer this, let us compare it to the magnitude of RAM-saving that the original DFTL enabled. Consider the minimal amount of RAM needed to store a pure RAM-based mapping where all mapping entries are in RAM: $y = K \cdot B \cdot 4$ bytes assuming 4 bytes per entry. Compare this to DFTL, where under Lazy Gecko, we need at least (($x = K \cdot B \cdot) / 8$ bytes). DFTL under Lazy Gecko allows reducing the RAM print a factor of only up to $x / y = 1 / 32$, a $97\%$ improvement. Logarithmic Gecko is capable of reducing this RAM footprint by a further $97\%$ on top. Thus, the magnitude of the reduction in RAM-consumption of moving from Lazy Gecko to Logarithmic Gecko is equivalent to the magnitude in RAM-reduction that DFTL enabled on the first place. 


In exchange for the lower RAM requirements, Logarithmic Gecko introduces some IO overheads relative to Lazy Gecko, which we capture in table \ref{tab:IO}. Let us start with the cost of an application write, which involves an insertion into the LSM-tree buffer. It is known that the cost of an insertion into an LSM-tree is $O(T/D \log_T(N/D))$, where N is the number of entries in the tree, D is the number of entries that fit into one page, and T is the size ratio between adjacent levels in the tree \cite{fractal-vs-lsm, lsm-tree}. In the case of Logarithmic Gecko's LSM-tree, N is equal to the number of blocks in the system K, and D is roughly equal to $P / B$, the number of block bitmaps that fit into one flash page. Thus, the cost of an application write in lazy gecko is $\frac{T \cdot B}{P} log_T(\frac{K \cdot B}{P}))$. Note that this expression is much lower than 1. In our experiments, the contribution of the LSM-tree to write-amplification was not greater than $3\%$. 

Let us now consider additional overheads introduced due to the two schemes during garbage-collection. Both Logarithmic Gecko and Lazy Gecko involve a cost of 1 flash read and 1 flash write per garbage-collection operation from and reading and rewriting a page from the reverse map. Logarithmic Gecko is associated with an additional cost for searching the LSM-tree to reconstruct the block validity bitmap for a candidate data block. In the worst case, each level of the tree must be searched, and searching each level involves at most 1 IO. Thus, the worst case cost is the number of levels in the tree: $log_T(\frac{K \cdot B}{P})$. 


\section{Evaluation}

\begin{figure}
\includegraphics[scale=0.25]{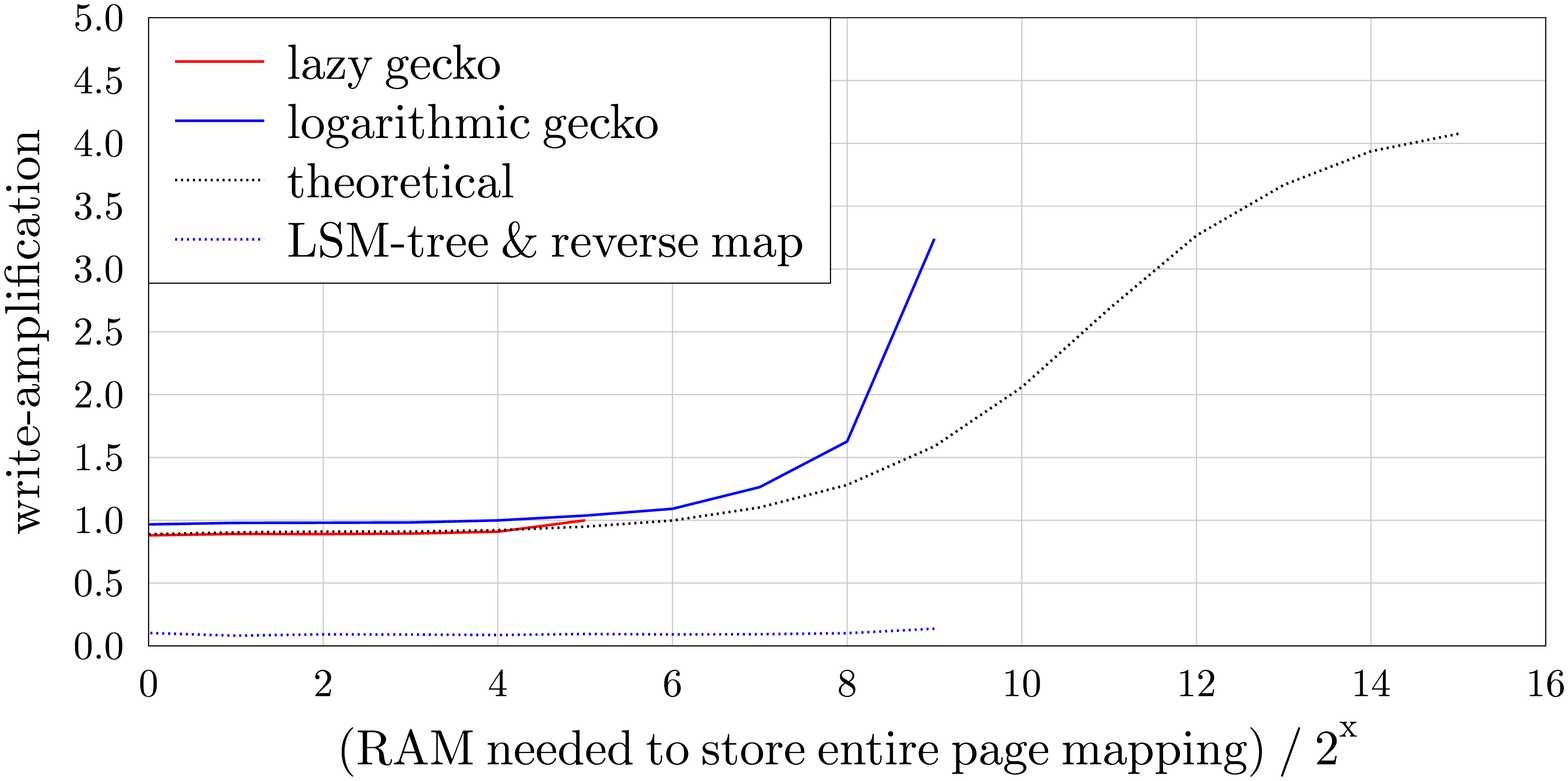}
\caption{Impact on write-amplification for different GC schemes as we decrease RAM}
 \label{fig:overall}
\end{figure}

We used the SSD simulator EagleTree \cite{EagleTree} to simulate an SSD similar to the Intel 525 series. It has the same features as in Table \ref{tab:terms}. OP was set to $30\%$. In our experiment, we varied the amount of RAM given to the SSD. We experimented with Lazy Gecko, Logarithmic Gecko, and a theoretical implementation of Lazy Gecko where all available RAM is used on the CMT. The workload we used consisted of uniformly randomly distributed writes across the logical address space. In Figure \ref{fig:overall}, we see how the different schemes allow the simulated SSD to scale. We start with an amount of RAM sufficient to store the entire page mapping in CMT, and halve it each time. We measure write-amplification. 

The theoretical implementation of Lazy Gecko increases in write-amplification as we decrease RAM due to more evictions from CMT but eventually levels off as we approach the theoretical worst case where each page write involves 1 eviction, so write-amplification is essentially doubled. Lazy Gecko performs as well as the theoretical optimal implementation, but it can't scale as the amount of RAM decreases. Logarithmic Gecko is able to scale to far lower levels of RAM than Lazy Gecko. Note that a bigger SSD such as the Micron P420m would be able to scale to even lower levels of RAM relative to Lazy Gecko, as we saw in Table \ref{tab:RAM}. The dotted blue line shows the contribution of the LSM-tree writes and reverse map writes to write-amplification for logarithmic Gecko, which is relatively low. Garbage-collection and evictions from CMT constitute the bulk of write-amplification. 

We also examined the impact of Logarithmic Gecko on read-amplification, or the number of internal reads that take place for each application write. The factors that contribute to read-amplification are (1) garbage-collection reads, (2) LSM-tree lookups and (3) mapping reads (both direct and reverse maps). Of these 3 factors in our experiment, contribution of the LSM-tree lookups to read-amplification ranges from $12\%$ when RAM is plentiful down to $3\%$ when RAM is scarce. This point is that the overheads of Logarithmic Gecko are just a small fraction of the overheads we would have had to pay anyway due to mapping reads and garbage-collection reads. 

\section{Conclusion}

We introduced the problem of maintaining the metadata needed to perform victim-selection and live-page-identification for garbage-collection in the context of flash-resident page-associative schemes. Lazy Gecko was introduced to solve this problem. It entails a modest IO overhead, and it requires storing a relatively large bitmap in RAM. We showed that this bitmap may introduce a scalability issue, and introduced Logarithmic Gecko, which stores this bitmap in flash as an LSM-tree. Logarithmic Gecko is able to scale to far lower quantities of RAM relative to the size of the SSD, and it only introduces a modest IO overhead due to maintaining and querying the LSM-tree.

\balance

\bibliographystyle{abbrv}
\bibliography{bib}  

\begin{thebibliography}{10}

\bibitem{tradeoff}
N.~Agrawal et~al.
\newblock Design tradeoffs for {SSD} performance.
\newblock In {\em USENIX}, pages 57--70, 2008.

\bibitem{assar}
M.~Assar, S.~Nemazie, and P.~Estakhri.
\newblock Flash memory mass storage architecture incorporation wear leveling
  technique, Dec.~26 1995.
\newblock US Patent 5,479,638.

\bibitem{EagleTree}
N.~Dayan, M.~K. Svendsen, M.~Bj{\o}rling, P.~Bonnet, and L.~Bouganim.
\newblock Eagletree: Exploring the design space of {SSD}-based algorithms.
\newblock {\em Proc. VLDB Endow.}, pages 1290--1293, Aug. 2013.

\bibitem{Desnoyers}
P.~Desnoyers.
\newblock Analytic modeling of {SSD} write performance.
\newblock In {\em SYSTOR}, pages 12:1--12:10, 2012.

\bibitem{bleak}
L.~M. Grupp, J.~D. Davis, and S.~Swanson.
\newblock The bleak future of {NAND} flash memory.
\newblock In {\em Proceedings of the 10th USENIX Conference on File and Storage
  Technologies}, FAST'12, pages 2--2, Berkeley, CA, USA, 2012. USENIX
  Association.

\bibitem{dftl}
A.~Gupta, Y.~Kim, and B.~Urgaonkar.
\newblock {DFTL}: A flash translation layer employing demand-based selective
  caching of page-level address mappings.
\newblock In {\em ASPLOS}, pages 229--240, 2009.

\bibitem{truth}
I.~Jim~Cooke, Micron~Technology.
\newblock The inconvenient truths of {NAND} flash memory.
\newblock In {\em Flash Memory Summit}, 2007.

\bibitem{ffs}
A.~Kawaguchi, S.~Nishioka, and H.~Motoda.
\newblock A flash-memory based file system.
\newblock In {\em USENIX}, pages 13--13, 1995.

\bibitem{BAST}
J.~Kim, J.~M. Kim, S.~H. Noh, S.~L. Min, and Y.~Cho.
\newblock A space-efficient flash translation layer for compactflash systems.
\newblock {\em IEEE Trans. on Consum. Electron.}, 48(2):366--375, May 2002.

\bibitem{fractal-vs-lsm}
B.~C. Kuszmaul.
\newblock A comparison of fractal trees to log-structured merge ({LSM}) trees.
\newblock {\em White Paper}, Apr. 2014.

\bibitem{F2FS}
C.~Lee, D.~Sim, J.~Hwang, and S.~Cho.
\newblock {F2FS}: A new file system for flash storage.
\newblock In {\em 13th USENIX Conference on File and Storage Technologies (FAST
  15)}, pages 273--286, Santa Clara, CA, 2015. USENIX Association.

\bibitem{FAST}
S.-W. Lee, W.-K. Choi, and D.-J. Park.
\newblock {FAST}: An efficient flash translation layer for flash memory.
\newblock In {\em EUC Workshops}, pages 879--887, 2006.

\bibitem{LazyFTL}
D.~Ma, J.~Feng, and G.~Li.
\newblock {LazyFTL}: A page-level flash translation layer optimized for {NAND}
  flash memory.
\newblock In {\em Proceedings of the 2011 ACM SIGMOD International Conference
  on Management of Data}, SIGMOD '11, pages 1--12, New York, NY, USA, 2011.
  ACM.

\bibitem{survey}
D.~Ma, J.~Feng, and G.~Li.
\newblock A survey of address translation technologies for flash memories.
\newblock {\em ACM Comput. Surv.}, 46(3):36:1--36:39, Jan. 2014.

\bibitem{lsm-tree}
P.~O'Neil, E.~Cheng, D.~Gawlick, and E.~O'Neil.
\newblock The log-structured merge-tree ({LSM}-tree).
\newblock {\em Acta Inf.}, 33(4):351--385, June 1996.

\bibitem{sast}
C.~Park, W.~Cheon, J.~Kang, K.~Roh, W.~Cho, and J.-S. Kim.
\newblock A reconfigurable {FTL} (flash translation layer) architecture for
  {NAND} flash-based applications.
\newblock {\em ACM Trans. Embed. Comput. Syst.}, 7(4):38:1--38:23, Aug. 2008.

\bibitem{consumer}
O.~T. Robert~Sykes.
\newblock Important differences between consumer and enterprise flash
  architectures.
\newblock Flash Memory Summit, Santa Clara, CA., 2013.

\bibitem{stoica}
R.~Stoica and A.~Ailamaki.
\newblock Improving flash write performance by using update frequency.
\newblock {\em Proc. VLDB Endow.}, pages 733--744, July 2013.

\bibitem{envy}
M.~Wu and W.~Zwaenepoel.
\newblock envy: A non-volatile, main memory storage system.
\newblock In {\em ASPLOS VI}, pages 86--97, 1994.

\end{thebibliography}

\end{document}